\documentclass[aps,pre,preprint,english]{revtex4-1}
\usepackage[latin9]{inputenc}
\usepackage{color}
\usepackage{babel}
\usepackage{mathrsfs}
\usepackage{amsmath}
\usepackage{amssymb}
\usepackage{graphicx}
\usepackage[unicode=true,
 bookmarks=false,
 breaklinks=false,pdfborder={0 0 1},backref=section,colorlinks=false]
 {hyperref}

\makeatletter
\@ifundefined{textcolor}{}
{%
 \definecolor{BLACK}{gray}{0}
 \definecolor{WHITE}{gray}{1}
 \definecolor{RED}{rgb}{1,0,0}
 \definecolor{GREEN}{rgb}{0,1,0}
 \definecolor{BLUE}{rgb}{0,0,1}
 \definecolor{CYAN}{cmyk}{1,0,0,0}
 \definecolor{MAGENTA}{cmyk}{0,1,0,0}
 \definecolor{YELLOW}{cmyk}{0,0,1,0}
}

\usepackage{babel}

\usepackage{changes}
\usepackage{float}
\usepackage{wasysym}
\usepackage{epstopdf}
\usepackage{float}
\usepackage{multirow}
\usepackage{scrextend}
\usepackage{epstopdf}
\usepackage{caption}
\usepackage{physics}
\usepackage{mathrsfs}
\usepackage{soul}
\usepackage{chemfig}

\usepackage{ulem}
\usepackage{soul}

\usepackage{ulem}
\usepackage{soul}
\usepackage{amsfonts}
\definechangesauthor[name=KLS, color=blue]{KLS}
\definechangesauthor[name=Tarun, color=orange]{TG}
\makeatother

\begin{document}

\title{ Effects of disorder on polaritonic and dark states in a cavity using the disordered Tavis-Cummings model}

\author{Tarun Gera and K.L. Sebastian\\Department of Inorganic and Physical Chemistry\\Indian Institute of Science\\ Bangalore 560012\\India}

\begin{abstract}
We consider molecules confined to a microcavity whose dimensions are such that an excitation of the molecule is nearly resonant with a cavity mode.   We investigate the situation where the excitation energies of the molecules are randomly distributed with a mean value of $\epsilon_a$ and variance $\sigma$.     For this case, we find a solution that approaches the exact result for large values of the number density $\mathscr{N}$ of the molecules.    We find the conditions for the existence of the polaritonic states, as well as expressions for their energies.  The polaritonic states are quite stable against disorder.  Analytical results are verified by comparison with simulations. When $\epsilon_a$ is equal to that of the cavity state $\epsilon_c$ (on resonance) the gap between the two polaritonic states is found to increase with disorder, the increase being equal to $2 \frac{\sigma^2}{\sqrt{\mathscr{N}}|\tilde{V}|}$ where $\tilde{V}$ is the coupling of a molecular excitation to the cavity state.  An analytic expression is found for the disorder induced width of the polaritonic peak. 
We results for various densities of states,  and the absorption spectrum.   The dark states that exist in the case $\sigma=0$  turn ``grey" in presence of disorder with their contribution to the absorption increasing with  $\sigma$. We analyze the effect of including lifetimes of the cavity and molecular states and find that in the strong coupling regime, the width of the polaritonic peaks is dominated by the lifetime effect and that disorder plays almost no role, if the Rabi splitting is sufficiently large.   We also consider the case where there is (a) orientational disorder as well as (b) spatial variation of the cavity field and find that they effectively amount to a renormlisation of the coupling. 
\end{abstract}

\maketitle

\section{Introduction}

In the recent past, states of molecules confined in a micro-cavity
have attracted much attention of chemists. The reason for this is
that such confinement can lead to very interesing modifications of
transport properties (energy and electron transfer reactions) and
chemical reactivity. These happen when an excited state of the molecule
is almost resonant with a cavity mode. In the strong coupling limit,
modified chemical reactivity \cite{Thomas2016,Thomas2019,Hirai2020,Peters2019,Munkhbat2018},
site selectivity \cite{Thomas2019}, altered enzyme activities \cite{Vergauwe2019}
and enhanced rate of energy transfer \cite{Zhong2016,Zhong2017,Coles2014,Akulov2018}
are some of the processes that have been reported. These applications
of light-matter interactions in chemistry that have been 
reported confirms the presence of hybrid light-matter states, for organic
dye molecules at room temperature in different states of matter confined
in a cavity. The Ebbesen group has been working on such systems and
has reported \cite{Thomas2016,Thomas2019,Ebbesen2016,Zhong2016,Zhong2017,Vergauwe2019,Hutchison2012,Orgiu2015}
many studies with various applications. Other research groups also
have recently published  results \cite{Hirai2020,Peters2019,Pietron2019,Munkhbat2018,Coles2014,Akulov2018,Casey2016}
 showing modified chemistry in such cavities. One of
the phenomena which is of great interest is the modification in chemical
reactivity, which can be achieved by making the cavity mode resonate
with different vibrational modes of the molecules. Many theoretical
models have been developed to understand this modified chemistry inside
the cavity. Transition state theory has been used to explain this
phenomenon but has not yet been able to provide a satisfactory explanation  \cite{Galego2019,Zhdanov2020,Jorge2020}.
This shows the need of continuous exploration and development of theories
which could explain the experimental results. The review articles
\cite{Baranov2018,Flick2018,Herrera2020,Herrera2018,Hertzog2019,Wang2021,Ribeiro2018,Feist2018}
compile the past developments, current status and unanswered questions
in this rapidly developing area of research.

When molecules are confined inside a cavity, there will be energy
transfer between excited states of the molecules and the cavity mode
of the confined electromagnetic field. If the rate of this energy
exchange is fast enough polaritonic modes are formed. For $N$ molecules
confined to the cavity that couple to the cavity mode through their
transition dipoles, the coupling leads to two polaritonic modes and
$N-1$ other states. In the case where all the molecules have identical
excitation energies, these $(N-1)$ modes are completely decoupled
from the electromagnetic field and hence cannot be excited by radiation.
Consequently, they are referred to as the ``dark modes". It is the formation
of these mixed light+matter states that leads to the new phenomena. The
polaritonic states have been predicted to lead to enhanced electron
transfer rates for molecules confined in the cavity \cite{Herrera2016}.
A very interesting experimental demonstration is the energy transfer mediated by polaritons \cite{Zhong2016,Coles2014}.

For molecules put inside a cavity there would be disorder in the energy
levels due to various reasons \cite{Scholes2020}. The causes  may be inhomogeneities in the material, different levels of aggregation,
dispersal by solvent or solvent fluctuations. It is well known that
in general, transport properties are strongly affected by the presence
of disorder. The reason for this is that disorder would in general
cause the states to be localized (Anderson localization) rather than
delocalized. For models in which one has nearest neighbour hopping,
even a small disorder leads to localization in $1D$ (dimension) and
$2D$. In $3D$ when the disorder exceeds a critical value the system would undergo
metal-insulator transition. Many aspects of disordered systems confined
to a cavity have been investigated, mostly by numerical methods by
Schachenmayer et. al. \cite{Botzung2020}. They consider the effect of the cavity on the transport on disordered systems, with nearest neighbor hopping.  When the disorder is large
enough, the states are localized and they investigate the effect of
the cavity on these localized states. They find that the dark states
are converted into ``grey states''. Further, increasing the coupling
to the cavity increases delocalization. The states can be spread over
multiple sites, which are not neighbors in space, but are in energy.
They find that these ``semi-localized'' states obey a semi-Poisson
statistics in the energy level spacings. Also, the coupling through
cavity modes is non-local and hence the effects are not dependent
on the dimensionality of the system. The semi-localized states are
found to be responsible for diffusive-like dynamics. It has been suggested
by Yuen-Zhou et al. \cite{Du2021}that the semi-localized states contribute
significantly to vibrational cooling, which leads to significant enhancement
of rate of electron transfer. In a new interesting work, Celardo et
al. \cite{Celardo2021} explore the long range hopping where transport
efficiency at first decreases, later increases and finally becomes
invariant as disorder increases.\\

Most of the investigations do not consider the position or orientational dependence of the coupling of the molecule with the cavity mode, as they assume the coupling to be a constant.  In addition, the inhomogeneities of the surrounding medium will affect the excitation energies, so that their values would be distributed randomly. Such disorder has been investigated previously in an interesting paper by Houdr$\acute{e}$ et. al. \cite{Houdre1996}.  Our analysis goes beyond theirs, in ways that are made clear in the Section \ref{Results} of this paper.    In the present work we study the effect of these disorders on the polaritonic
and the dark states. Using the Tavis-Cummings Hamiltonian \cite{Tavis1968,E1969}
with added disorder, we analytically solve for the Green's function
for the system in the limit $N>>1$. The density of states and absorption
cross-section  also are analytically expressed in terms of the Green's
function.  In section {\ref{Sec:Model1}} we consider the model where the
coupling of each molecule to the cavity state is considered to be
the same, but their excitation energies are randomly distributed.  Having illustrated our approach with this simple model,  in section {\ref{Sec:Model2} we discuss the more general case  in which 
the coupling depends on the position as well as the orientation of
the transition dipole moment of the molecule with respect to the cavity field. To see the effects of disorder clearly we have taken the lifetime of the excited molecular state to be very long. We also discuss the inclusion of lifetime effects.  It is found that the polaritons have lifetimes which are weighted averages of the lifetimes of the cavity and the molecular states. Further, on resonance the disorder effects on the line-broadening are negligible  if $\Omega>>\frac{\gamma_a+\gamma_c}{2}>>\sigma$, where $\Omega$ is the rabi splitting and $\gamma_a$ and $\gamma_c$ are the lifetimes of the molecular and cavity states.
\section{The Hamiltonian and the Green's operator}

In the Tavis-Cummings Hamiltonian \cite{Tavis1968,E1969} with disorder
in the excitation energies, we take the excitation energy $\epsilon_{i}$
of the $i^{th}$ molecule to be 
\begin{equation}
\epsilon_{i}=\epsilon_{a}+\xi_{i},\label{Eq.epsilon1}
\end{equation}
where $\xi_{i}$ is a random variable.

The total Hamiltonian may then be written as 
\begin{equation}
\hat{H}=\epsilon_{c}|c\rangle\langle c|+\sum_{i}^{N}(\epsilon_{a}+\xi_{i})|a_{i}\rangle\langle a_{i}|+\sum_{i}^{N}(V_{i}|c\rangle\langle a_{i}|+V_{i}^{*}|a_{i}\rangle\langle c|).
\end{equation}
$|c\rangle$ denotes the state where the cavity mode is excited to
its first excited state and it has an  energy $\epsilon_{c}$. $|a_{i}\rangle$
denotes the state in which the $i^{th}$ molecule is excited with all the other molecules in their ground states. The
coupling constant of the cavity mode to this  excitation is
denoted by $V_{i}$ which is given by 
\begin{equation}
V_{i}=-\boldsymbol{E}(z_i).\boldsymbol{{\mu}}_i=-\sqrt{\frac{\epsilon_{c}}{2\epsilon_{o}{\cal V}}}\mu_{eg}\sin(k_{c}z_{i})\cos(\theta_{i}),\label{cc}
\end{equation}
where ${\cal V}$ is the volume of the cavity and $\boldsymbol{{\mu}}_i$ is transition dipole moment of the molecular excitation, having  the magnitude  $\mu_{eg}$.  The cavity field wave-vector and the permittivity of free space are denoted as  $k_c$ and $\epsilon_{o}$ respectively.  The term $\sin(k_{c}z_{i})$ accounts for the spatial variation of the cavity field and $\theta_{i}$ is the angle between the transition dipole moment vector and the electric field vector.
 We use the Green's operator to analyze the problem
following the techniques of the classic paper of Anderson on the impurity
problem \cite{Anderson1961}. See also the book by Mattuck \cite{Mattuck2012}.
The Green's operator for the Hamiltonian $\hat{H}$ is defined by
$\hat{G}(\omega)=(\omega-\hat{H})^{-1}$, where $\omega$ is a complex
variable. We shall calculate the matrix element $G_{ij}(\omega)$
of $\hat{G}(\omega)$ defined by 
\begin{equation}
G_{ij}(\omega)=\langle i|\frac{1}{\omega-\hat{H}}|j\rangle,
\end{equation}
where $|i\rangle$ and $|j\rangle$ are any two arbitrary states. $G_{ij}(\omega)$
has the spectral representation 
\begin{equation}
G_{ij}(\omega)=\sum_{m}\frac{\langle i|\varepsilon_{m}\rangle\langle\varepsilon_{m}|j\rangle}{\omega-\varepsilon_{m}},
\end{equation}
where $|\varepsilon_{m}\rangle$ is an eigenfunction of the Hamiltonian
operator having the eigenvalue $\varepsilon_{m}$, satisfying $\hat{H}|\varepsilon_{m}\rangle=\varepsilon_{m}|\varepsilon_{m}\rangle.$
Note that $m$ runs from $1$ to $(N+1)$. From this it is clear that
any discrete eigenvalue of $\hat{H}$ appears as a simple pole of
$G_{ij}(\omega)$ and that its residue at the pole at $\varepsilon_{m}$
is $\langle i|\varepsilon_{m}\rangle\langle\varepsilon_{m}|j\rangle$,
giving us an idea of the amount of contribution that the $m^{th}$
eigenstate makes to $|i\rangle$ and $|j\rangle$. We now proceed
to determine the matrix elements $G_{ij}(\omega)$.

The Hamiltonian may be written in the matrix form as 
\begin{equation}
\underline{\underline{H}}=\left[\begin{array}{c|ccccccc}
\epsilon_{c}\; & \;V_{1}\; & \;V_{2}\; & \;.\; & \;.\; & \;.\; & \;.\; & \;V_{N}\;\\
\hline V_{1}^{*} & \epsilon_{1} & 0 & . & . & . & . & 0\\
V_{2}^{*} & 0 & \epsilon_{2} & . & . & . & . & 0\\
. & . & . & . & . & . & . & .\\
V_{N}^{*} & 0 & 0 & . & . & . & . & \epsilon_{N}
\end{array}\right],
\end{equation}
which has the partitioned form 
\begin{equation}
\underline{\underline{H}}=\left[\begin{array}{c|c}
\epsilon_{c} & \underline{V}\\
\hline \underline{V}^{\dagger} & \underline{\underline{\epsilon}}_{M}
\end{array}\right].\label{Eq.H1}
\end{equation}
$M$ stands for molecules and $\underline{V}=[V_{1},V_{2},...,V_{N}]$
is a row matrix having dimensions $1\times N$. $\underline{\underline{\epsilon}}_{M}$
is an $N\times N$ diagonal matrix of the excitation energies: 
\begin{equation}
\left(\underline{\underline{\epsilon}}_{M}\right)_{ij}=\epsilon_{i}\delta_{ij}.
\end{equation}
The matrix corresponding to $\hat{G}(\omega)$ operator is 
\begin{equation}
\underline{\underline{G}}=\left(\omega{\underline{\underline{I}}_{N+1}}-\underline{\underline{H}}\right)^{-1}.\label{Eq.G1}
\end{equation}
$\underline{\underline{I}}_{N+1} $ is the $ (N+1)$ dimensional identity matrix. $\underline{\underline{G}}$ too can be written in the same partitioned
form as in Eq. (\ref{Eq.H1}): 
\begin{equation}
\underline{\underline{G}}=\left[\begin{array}{c|c}
G_{cc} & \underline{G}_{cM}\\
\hline \underline{G}_{cM}^{\dagger} & \underline{\underline{G}}_{M}
\end{array}\right].\label{Eq.G2}
\end{equation}
Eq. (\ref{Eq.G1}) can be rearranged to get 
\begin{equation}
(\omega\underline{\underline{I}}_{N+1}-\underline{\underline{H}})\underline{\underline{G}}=\underline{\underline{I}}_{N+1}.
\end{equation}
 Using
Eq. (\ref{Eq.H1}), Eq. (\ref{Eq.G1}) and Eq. (\ref{Eq.G2}) we get
\begin{equation}
\begin{pmatrix}\omega-\epsilon_{c} & -\underline{V}\\
-\underline{V}^{\dagger} & \omega\underline{\underline{I}}-\underline{\underline{\epsilon}}_{M}
\end{pmatrix}\begin{pmatrix}G_{cc} & \underline{G}_{cM}\\
\underline{G}_{cM}^{\dagger} & \underline{\underline{G}}_{M}
\end{pmatrix}=\begin{pmatrix}1 & 0\\
0 & \underline{\underline{I}}
\end{pmatrix}.\label{Eq.HG-1}
\end{equation}
In the above $\underline{\underline{I}}$ is the $N\times N$ identity
matrix. Eq. (\ref{Eq.HG-1}) is equivalent to the following four equations:
\begin{eqnarray}
(\omega-\epsilon_{c})G_{cc}-\underline{V}\,\underline{G}_{cM}^{\dagger} & = & 1\label{Eqset.1}\\
(\omega-\epsilon_{c})\underline{G}_{cM}-\underline{V}\,\underline{\underline{G}}_{M} & = & 0\\
-\underline{V}^{\dagger}G_{cc}+(\omega\underline{\underline{I}}-\underline{\underline{\epsilon}}_{M})\underline{G}_{cM}^{\dagger} & = & 0\label{Eqset.2}\\
-\underline{V}^{\dagger}\underline{G}_{cM}+(\omega\underline{\underline{I}}-\underline{\underline{\epsilon}}_{M})\underline{\underline{G}}_{M} & = & \underline{\underline{I}}.\label{Eqset.4}
\end{eqnarray}
Solving Eq. (\ref{Eqset.2}) for $\underline{G}_{cM}^{\dagger}$ gives
\begin{equation}
\underline{G}_{cM}^{\dagger}=(\omega\underline{\underline{I}}-\underline{\underline{\epsilon}}_{M})^{-1}\underline{V}^{\dagger}G_{cc}.\label{Eq.Gcmdaggger}
\end{equation}
Using Eq. (\ref{Eq.Gcmdaggger}) in Eq. (\ref{Eqset.1}) and solving
for $G_{cc}$ leads to 
\begin{equation}
G_{cc}=\left[\omega-\epsilon_{c}-\Sigma(\omega)\right]^{-1},\label{Eq.Gcc1}
\end{equation}
with the self energy $\Sigma(\omega)$ defined by 
\begin{equation}
\Sigma(\omega)=\sum_{i}\frac{V_{i}^{*}V_{i}}{\omega-\epsilon_{i}}.\label{Eq.SE-1}
\end{equation}
We define the imaginary and real parts of the self energy by: 
\[
\Sigma_{I}(\omega)=-\pi\sum_{i}|V_{i}|^{2}\delta(\omega-\epsilon_{i})
\]
and 
\[
\Sigma_{R}(\omega)=-\frac{1}{\pi}{\cal P}\int_{-\infty}^{\infty}d\epsilon\frac{\Sigma_{I}(\epsilon)}{\omega-\epsilon},
\]
where ${\cal P}$ stands for the principal value of the integral.
It is useful to introduce the density of states for the disordered
molecular states by 
\begin{equation}
\rho_{d}(\xi)=\frac{1}{N}\sum_{i=1}^{N}\delta(\xi-\xi_{i}).
\end{equation}
In the following we first consider the simplest possible
model where $V_{i}=V,$ a constant. For this model, all the calculations
are reported in detail. After that we also discuss more general cases
in section {\ref{Sec:Model2}}.

\section{Model I: $V_{i}=V$}\label{Sec:Model1}

We note that the coupling $V\propto\frac{1}{\sqrt{{\cal {V}}}}$, where ${\cal {V}}$
is the volume of the cavity. We define $\mathscr{N}$ the number density of molecules by  
 $\mathscr{N} =N/{\cal {V}}$ is the number density of molecules and volume independent coupling  $\tilde{V}=V\sqrt{{\cal {V}}}$.  \\

 We use the notation $\omega^{+}=\omega+i\eta$
when $\omega$ is real and $\eta$ is taken to be an infinitestimal
positive number. We then have 
\begin{equation}
\Sigma(\omega^{+})=\mathscr{N}\tilde{V}^{2}\int_{-\infty}^{\infty}d\xi\frac{\rho_{d}(\xi)}{\omega+i\eta-\epsilon_{a}-\xi}.
\end{equation}
Separating out the real and imaginary parts of $\Sigma(\omega^{+})$
gives 
\begin{equation}
\Sigma(\omega^{+})=\mathscr{N}\tilde{V}^{2}{\cal {P}}\int_{-\infty}^{\infty}d\xi\frac{\rho_{d}(\xi)}{\omega-\epsilon_{a}-\xi}-i\pi\mathscr{N}\tilde{V}^{2}\int_{-\infty}^{\infty}d\xi\rho_{d}(\xi)\delta(\omega-\epsilon_{a}-\xi),
\end{equation}
where ${\cal {P}}$ denotes the principal value of the integral.

In this simplest model with $V_{i}=V$, the coupling with the cavity
field is the same for all molecules. Further we shall take $\xi_{i}$
to be identically distributed Gaussian random variables with the probability
distribution 
\begin{equation}
P(\xi)=\frac{1}{\sqrt{2\pi}\sigma}e^{\frac{-\xi^{2}}{2\sigma^{2}}}.
\end{equation}
In principle, one has to average over different realisations of the
random quantities $\xi_i$. However when ${\mathscr N}$ is large, we expect $\rho_{d}(\omega)\rightarrow P(\xi).$
Hence in this limit one expects \emph{the ensemble averaged value}
of self energies to be given by: 
\begin{equation}
\Sigma_{I}(\omega)=\frac{\mathscr{N}\tilde{V}^{2}\pi}{\sqrt{2\pi}\sigma}e^{-\frac{(\omega-\epsilon_{a})^{2}}{2\sigma^{2}}}
\end{equation}
and 
\begin{equation}
\Sigma_{R}(\omega)=\frac{\sqrt{2}\mathscr{N}\tilde{V}^{2}}{\sigma}DawsonF\left(\frac{\omega-\epsilon_{a}}{\sqrt{2}\sigma}\right).
\end{equation}
$DawsonF(x)$ is the Dawson Integral (also referred to as Dawson function) \cite{DawsonMathworld},
defined by  
\begin{equation}
DawsonF(x)=e^{-x^{2}}\int_{0}^{x}dy\;e^{y^{2}}.
\end{equation}
Using these results in Eq. (\ref{Eq.Gcc1}) we get\\
\begin{equation}
G_{cc}(\omega)=\left[\omega-\epsilon_{c}-\frac{\sqrt{2}\mathscr{N}\tilde{V}^{2}}{\sigma}DawsonF\left(\frac{\omega-\epsilon_{a}}{\sqrt{2}\sigma}\right)+\frac{i\mathscr{N}\tilde{V}^{2}\pi}{\sqrt{2\pi}\sigma}e^{\frac{-(\omega-\epsilon_{a})^{2}}{2\sigma^{2}}}\right]^{-1}.
\end{equation}

\subsection{The Density of States}

We define the density of states for the cavity state by 
\begin{eqnarray}
\rho_{c}(\omega) & = & \sum_{m=1}^{N+1}\left|\langle c|m\rangle\right|^{2}\delta(\omega-\varepsilon_{m})\nonumber \\
 & = & -\frac{1}{\pi}Im\{G_{cc}(\omega^{+})\}.
\end{eqnarray}
$\rho_c(\omega)$ gives us an idea of the amount of participation of the cavity state in the $m^{th}$ eigenstate of the system. \\

 Now we calculate the matrix elements of Green's operator for the molecular
states. Eq. (\ref{Eqset.2}) can be solved for $\underline{G}_{cM}$
to get 
\begin{equation}
\underline{G}_{cM}=(\omega-\epsilon_{c})^{-1}\underline{V}\,\underline{\underline{G}}_{M}.
\end{equation}
Using this result in Eq. (\ref{Eqset.4}) gives 
\begin{equation}
\underline{\underline{G}}_{M}(\omega^{+})=\left[\underline{\underline{I}}\omega^{+}-\underline{\underline{\epsilon}}_{M}-\frac{\underline{V}^{\dagger}\underline{V}}{\omega^{+}-\epsilon_{c}}\right]^{-1}.\label{Eq.GM}
\end{equation}
The operator corresponding to the matrix $\underline{V}^{\dagger}\underline{V}$
can be written in the form $\mathscr{N}\tilde{V}^{2}\left|mol\right\rangle \left\langle mol\right|$
with 
\begin{equation}
\left|mol\right\rangle =\frac{1}{\sqrt{N}}\sum_{i=1}^{N}\left|i\right\rangle .\label{eq:|mol>-definition}
\end{equation}
Hence we can write the operator corresponding to the matrix $\underline{\underline{G}}_{M}(\omega^{+})$
as 
\begin{equation}
\hat{G}_{M}(\omega^{+})=\left[\omega^{+}-\hat{\epsilon}_{M}-\frac{\mathscr{N}\tilde{V}^{2}\left|mol\right\rangle \left\langle mol\right|}{\omega^{+}-\epsilon_{c}}\right]^{-1}.
\end{equation}
The operator $\hat{\epsilon}_{M}$ is defined by its matrix elements
$\langle i|\hat{\epsilon}_{M}|i'\rangle=\epsilon_{i}\;\delta_{ii'}$. Using the
operator identity $(\hat{A}-\hat{B})^{-1}=\hat{A}^{-1}+\hat{A}^{-1}\hat{B}(\hat{A}-\hat{B})^{-1}$ valid for any two operators $\hat{A}$ and $\hat{B}$,
we can write 
\begin{align}
\hat{G}_{M}(\omega^{+}) & =\left[\omega^{+}-\hat{\epsilon}_{M}\right]^{-1}+\left[\omega^{+}-\hat{\epsilon}_{M}\right]^{-1}\frac{\mathscr{N}\tilde{V}^{2}\left|mol\right\rangle \left\langle mol\right|}{\omega^{+}-\epsilon_{c}}\hat{G}_{M}(\omega^{+}).\label{eq:G_M-1}
\end{align}
From this we find $\langle mol|\hat{G}_{M}(\omega^{+})$ to obey
the equation 
\begin{align}
\langle mol|\hat{G}_{M}(\omega^{+}) & =\langle mol|\left[\omega^{+}-\hat{\epsilon}_{M}\right]^{-1}+\frac{\Sigma(\omega^{+})}{(\omega^{+}-\epsilon_{c})}\langle mol|\hat{G}_{M}(\omega^{+}).
\end{align}
Solving for $\langle mol|\hat{G}_{M}(\omega^{+})$ gives 
\begin{align*}
\langle mol|\hat{G}_{M}(\omega^{+})= & (\omega^{+}-\epsilon_{c})G_{cc}(\omega^{+})\langle mol|\left[\omega^{+}-\hat{\epsilon}_{M}\right]^{-1}.
\end{align*}
Using this back in Eq. (\ref{eq:G_M-1}) we get 
\begin{equation}
\hat{G}_{M}(\omega^{+})=\left[\omega^{+}-\hat{\epsilon}_{M}\right]^{-1}+\left[\omega^{+}-\hat{\epsilon}_{M}\right]^{-1}\mathscr{N}\tilde{V}^{2}\left|mol\right\rangle G_{cc}(\omega^{+})\langle mol|\left[\omega^{+}-\hat{\epsilon}_{M}\right]^{-1}.\label{eq:GM-2}
\end{equation}
Using the above, the matrix element $G_{mol,mol}(\omega^{+})=\left\langle mol\right|\hat{G}_{M}(\omega^{+})\left|mol\right\rangle $
can be calculated to be: 
\begin{eqnarray}
G_{mol,mol}(\omega^{+}) & = & \left\{ 1-\frac{\mathscr{N}\tilde{V}^{2}}{\omega^{+}-\epsilon_{c}}\left\langle mol\right|(\omega^{+}-\hat{\epsilon}_{M})^{-1}\left|mol\right\rangle \right\} ^{-1}\left\langle mol\right|(\omega^{+}-\hat{\epsilon}_{M})^{-1}\left|mol\right\rangle \\
 & = & (\omega^{+}-\epsilon_{c})G_{cc}(\omega^{+})\frac{1}{\mathscr{N}\tilde{V}^{2}}\Sigma(\omega^{+}).
\end{eqnarray}
From $G_{mol,mol}(\omega^{+})$ we can calculate the density of states
of $|mol\rangle:$
\begin{equation}
\rho_{mol}(\omega)=-\frac{1}{\pi}Im\{G_{mol,mol}(\omega^{+})\}.\label{eq:rho_mol}
\end{equation}

The total density of states may be defined by $\rho_{T}(\omega)=\sum_{m=1}^{N+1}\delta(\omega-\varepsilon_{m})$
which may be written as $\rho_{T}(\omega)=-\frac{1}{\pi}Im \left ( Tr{\hat{G}(\omega^{+}) } \right )$.
Using the complete set $|c\rangle,|1\rangle,|2\rangle,....|N\rangle$
to calculate the trace leads to 
\begin{eqnarray}
\rho_{T}(\omega) & = & \rho_{c}(\omega)-\sum_{i=1}^{N}\frac{1}{\pi}Im\{G_{ii}(\omega^{+})\},\label{Eq.rhot}
\end{eqnarray}
with $G_{ii}(\omega^{+})=\langle i|\hat{G}(\omega^{+})|i\rangle$.
$G_{ii}(\omega^{+})$ for $i\neq c$ may be easily evaluated using Eq.
(\ref{eq:GM-2}), to get 
\[
G_{ii}(\omega^{+})=(\omega^{+}-\epsilon_{i})^{-1}+V^{2}(\omega^{+}-\epsilon_{i})^{-2}G_{cc}(\omega^{+}),
\]
so that 
\begin{equation}
\sum_{i=1}^{N}G_{ii}(\omega^{+})=\sum_{i=1}^{N}(\omega^{+}-\epsilon_{i})^{-1}+V^{2}G_{cc}(\omega^{+})\sum_{i=1}^{N}(\omega^{+}-\epsilon_{i})^{-2}.\label{eq:TrG}
\end{equation}
A change in the total density of states for the molecules alone may
be defined by

\[
\Delta\rho_{M}(\omega)=-\frac{1}{\pi}Im\left(\sum_{i=1}^{N}G_{ii}(\omega^{+})\right)-\rho_{M}^{0}(\omega),
\]
where $\rho_{M}^{0}(\omega)$ is the density of molecular states in
the case where $\tilde{V}=0$. The above may be calculated using Eq.
(\ref{eq:TrG}) to be 
\begin{equation}
\Delta\rho_{M}(\omega)=\frac{1}{\pi}Im\left(\frac{\partial\Sigma(\omega^{+})}{\partial\omega}G_{cc}(\omega^{+})\right)\label{Eq.drhom}
\end{equation}
so that the change in the total density of states, due to the interaction
of the cavity mode with the molecules is 
\begin{equation}
\Delta\rho_{T}(\omega)=-\frac{1}{\pi}Im\{G_{cc}(\omega^{+})\}-\frac{1}{\pi}Im\{\frac{\partial\Sigma(\omega^{+})}{\partial\omega}G_{cc}(\omega^{+})\}.\label{eq:Delta-rho_T}
\end{equation}

\subsection{The Absorption Spectrum}

The Hamiltonian for the interaction of the system with radiation is
\[
\hat{H}_{int}(t)=-\sum_{i}\text{\ensuremath{\boldsymbol{E}_{i}}}(t).\boldsymbol{\hat{\mu}}_{i}.
\]
The $\boldsymbol{\hat{\mu}}_{i}$  operator can be written in terms in terms of transition dipole moment as:
\[
\boldsymbol{\hat{\mu}}_{i}=\boldsymbol{\mu_i} |g\rangle\langle i| + \boldsymbol{\mu_i^{*}} |i\rangle\langle g|
\]
where $\boldsymbol{\mu_{i}}=\langle i|\boldsymbol{\hat{\mu}}_{i}|g\rangle$. We can rewrite the Hamiltonian for interaction as: 
\begin{equation}
\hat{H}_{int}(t)=-\sum_{i}\boldsymbol{E}_{i}(t).\left(\boldsymbol{\mu_i} |g\rangle\langle i| + \boldsymbol{\mu_i^{*}} |i\rangle\langle g|\right).\label{Eq.Hint}
\end{equation}
$|g \rangle$ indicates the ground state in which all the molecules are in their ground states.   $|i\rangle$ is the site in which the $i^{th}$ molecule is excited.  $\boldsymbol{E}_{i}(t)$
is the harmonic electric field of the frequency $\omega$ that the
$i^{th}$ molecule experiences. In the spirit of the model that assumes
$V_{i}=V$ a constant, we first take the electric field to be the
same for all the molecules - i.e., neglect its space dependence so
that all the molecules experience the same electric field. Further,
for simplicity, we take all of them to be oriented in the direction
of the electric field. Under these assumptions,
\[
\hat{H}_{int}(t)=-\sum_{i}E(t)\left(\mu_{eg}|g\rangle\langle i|+\mu_{ge}^{*}|i\rangle\langle g|\right)
\]
where $\mu_{eg}=|\boldsymbol{\mu_i}|$ 
and $E(t)$ is the electric field in the direction of the transition
dipole of $i^{th}$ molecule. The absorption cross section $\alpha(\omega)$
for radiation of frequency $\omega$ is then  
\[
\alpha(\omega)=\frac{\pi\omega}{\epsilon_{o}c\hbar}\;\;\sum_{f}\left|\left\langle f\left|\hat{H}_{int}(0)\right|g\right\rangle \right|^{2}\delta(E_{f}-E_{g}-\hbar\omega).
\]
In the above, $E_g$ is the energy of the ground state.  $|f\rangle$ denotes a possible final state, having energy $E_{f}$. Following reference
\cite{schatz2012} we write this as the Fourier transform of a correlation
function: 
\[
\alpha(\omega)=\frac{\omega}{2\epsilon_{o}c\hbar}\int_{-\infty}^{\infty}dte^{i\omega t}\left\langle g\left|\hat{H}_{int}(t)\hat{H}_{int}\right|g\right\rangle, 
\]
where $\hat{H}_{int}(t)=e^{i\hat{H}t/\hbar}\hat{H}_{int}(0)e^{-i\hat{H}t/\hbar}$.
On using the expression for $H_{int}$ and simplifying we get
\begin{align}
\alpha(\omega) & =\frac{\omega}{2\epsilon_{o}c\hbar}\left|\boldsymbol{{\mu}}\right|^{2}\int_{-\infty}^{\infty}dte^{i\omega t}\sum_{i,j}\left\langle i\left|e^{i\hat{H}t/\hbar}\right|j\right\rangle \nonumber \\
 & =-\frac{\omega}{\epsilon_{o}c\hbar}N\left|\boldsymbol{{\mu}}\right|^{2}\;Im\left[G_{mol,mol}(\omega)\right].\label{eq:Eq.CS}
\end{align}
\section{Results}
\label{Results}
\subsection{Existence of polaritonic states}
We now analyze $G_{cc}(\omega)$ to see the effect of disorder on
the polaritonic states. At small disorder one expects the polaritonic
states to exist. However, as the disorder increases, one suspects
that the polaritonic states may disappear. For the sake of simplicity we consider only the case where $\epsilon_c=\epsilon_a$, and it is easy to extend the analysis to more general cases. In the following we make 
a quantitative analysis using the figures \ref{Fig1} and
\ref{Fig2}, which   differ in their values of ${\mathscr N}$. Both
show the plots of $\omega-\epsilon_{c}$, $\Sigma_{R}(\omega-\epsilon_{c})$
and $\Sigma_{I}(\omega-\epsilon_{c})$ against $\omega-\epsilon_{c}$.
A pole of $G_{cc}(\omega)$ occurs at a point where the
plot of $\omega-\epsilon_{c}$ intersects with that of $\Sigma_R(\omega-\epsilon_{c})$.  Fig. \ref{Fig1} is for large values of ${\mathscr N}$ and
in this case there are three points where this happens.  The intersection  at the origin has a large value for the imaginary part $\Sigma_I(\omega-\epsilon_c)$ implying that this a virtual state.  The other two intersections have only a small value of $\Sigma_I(\omega-\epsilon_c)$ indicating these are the long lived polaritonic states. 
 This figure is to be compared
with Fig. \ref{Fig2} which is for smaller values of ${\mathscr N}$ or
larger value of $\sigma$. In this case the polaritonic states do
not exist. Only the virtual state is there. The plot of $\Sigma_{R}(\omega-\epsilon_{c})$
has a slope of $\frac{\mathscr{N}\tilde{V}^{2}}{\sigma^{2}}$ at $\omega=\epsilon_{c}$.
Hence the two polaritonic poles can exist only if $\frac{\mathscr{N}\tilde{V}^{2}}{\sigma^{2}}>1$.
For suitable values of the parameters, it is also possible that only
one polaritonic state exists.

\begin{figure}[h]
\includegraphics[width=0.5\linewidth]{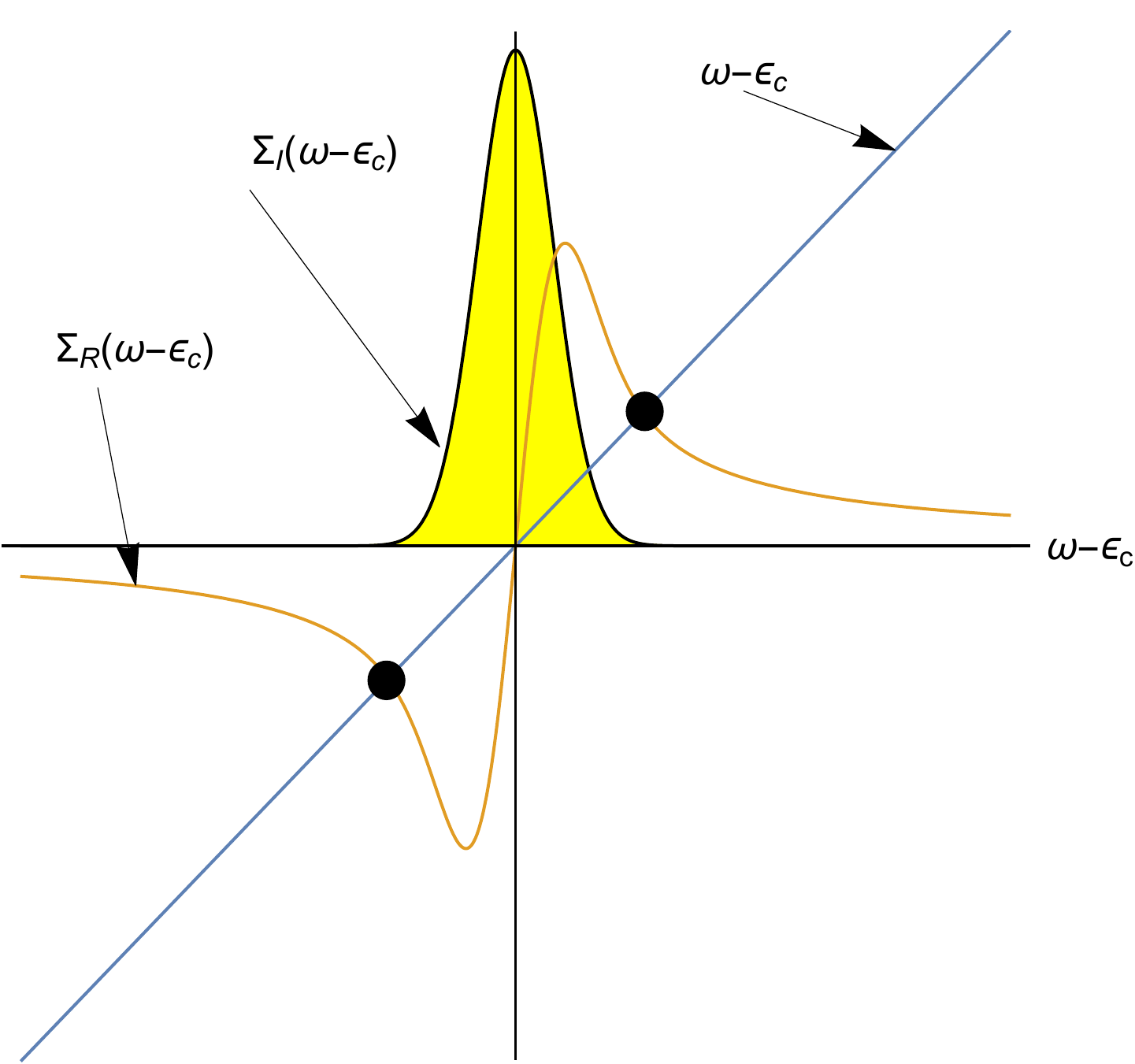}\caption{This figure takes $\epsilon_a$ to be equal to $\epsilon_c$. The plots of $\omega-\epsilon_{c},\;\Sigma_{R}(\omega-\epsilon_{c})$
and $\Sigma_{I}(\omega-\epsilon_{c})$ against $(\omega-\epsilon_{c}).$
The darkened small circles indicate points where $\omega-\epsilon_{c}$
intersects with $\Sigma_{R}(\omega-\epsilon_{c})$ leading to the
formation of two discrete poles. These correspond to the polaritonic states.
The intersection at the origin indicates a virtual state as the imaginary
part of the self energy $\Sigma_{I}(0)$ is large at this point.}
\label{Fig1} 
\end{figure}

\begin{figure}[h]
\includegraphics[width=0.5\linewidth]{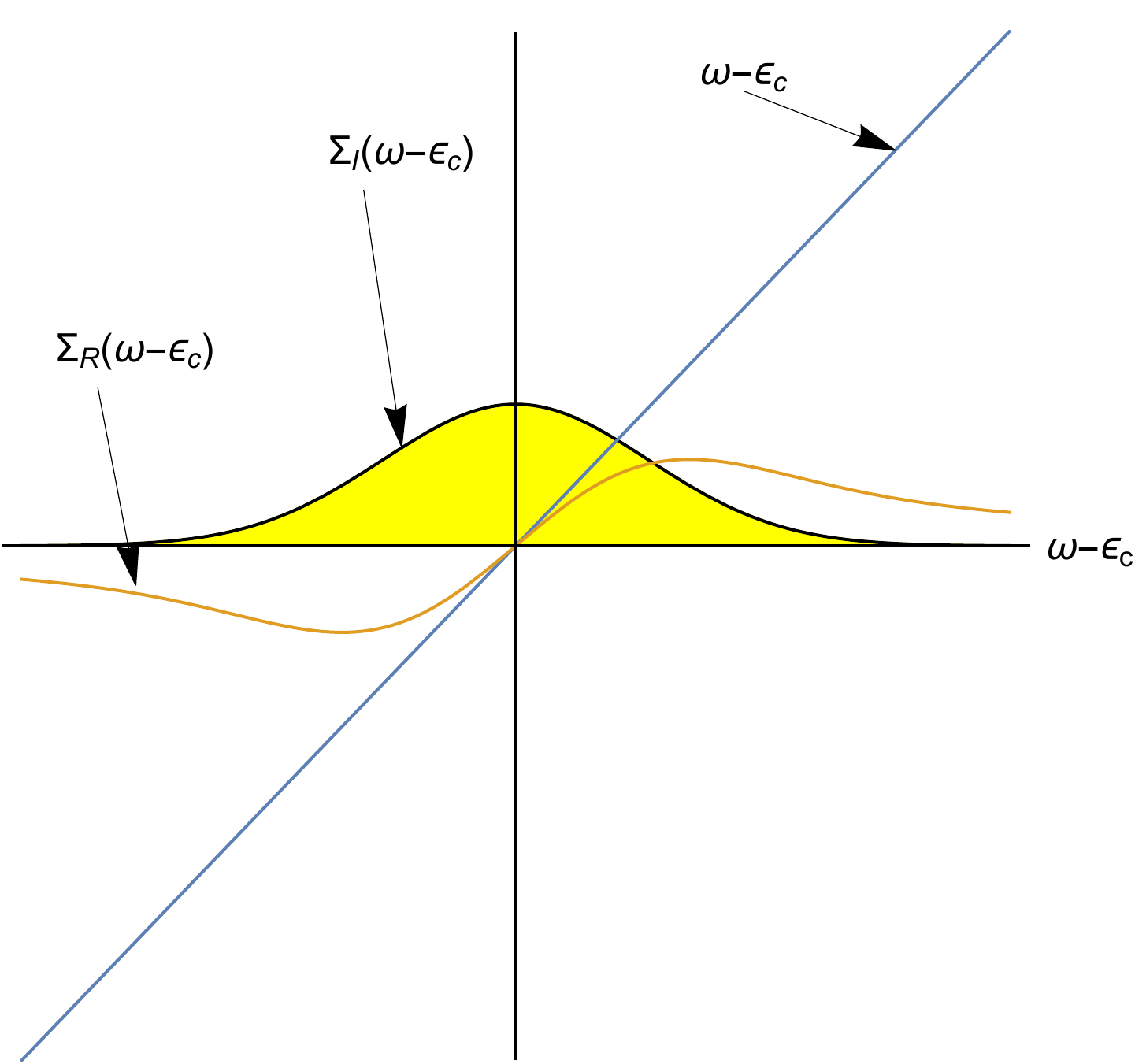} \caption{This figure takes $\epsilon_a$ to be equal to $\epsilon_c$.  The plots of $\omega-\epsilon_{c}$,\;$\Sigma_{R}(\omega-\epsilon_{c})$
and $\Sigma_{I}(\omega-\epsilon_{c})$ against $(\omega-\epsilon_{c})$
for increased disorder. There is only one point where $\omega-\epsilon_{c}$
intersects with $\Sigma_{R}(\omega-\epsilon_{c})$ and that is just
the virtual state. }
\label{Fig2} 
\end{figure}

We now make an analysis that is valid for arbitrary values of $\epsilon_a$ and $\epsilon_c$. We  consider the case where disorder is not large so that
the situation in Fig. \ref{Fig1} occurs and there are two poles for
$G_{cc}(\omega)$.  These poles occur for large values of $|\omega-\epsilon_{c}|$ and $|\omega-\epsilon_{a}|$ .
The Dawson Function has the asymptotic form 
\[
DawsonF(x)=\frac{1}{2x}+\frac{1}{4x^{3}}+\frac{3}{8x^{5}}+O\left(\frac{1}{x^{7}}\right)\;\;\;\text{for \ensuremath{|x|} large.}
\]
Using this to the lowest order (i.e. only the $1/x$ term is retained),
we find that for $\left|\frac{\omega-\epsilon_{a}}{\sqrt{2}\sigma}\right|>>1$,
$\frac{\sqrt{2}\mathscr{N}\tilde{V}^{2}}{\sigma}DawsonF\left(\frac{\omega-\epsilon_{a}}{\sqrt{2}\sigma}\right)\simeq\frac{\mathscr{N}\tilde{V}^{2}}{(\omega-\epsilon_{a})}$
and hence 
\[
G_{cc}(\omega)=\left(\omega-\epsilon_{c}-\frac{\mathscr{N}\tilde{V}^{2}}{(\omega-\epsilon_{a})}+i\frac{\mathscr{N}\tilde{V}^{2}\pi}{\sqrt{2\pi}\sigma}e^{\frac{-(\omega-\epsilon_{a})^{2}}{2\sigma^{2}}}\right)^{-1}.
\]
As $(\omega-\epsilon_{a})/(\sqrt{2}\sigma)>>1,$ the imaginary part
in the above is very small and may be neglected for a first analysis.
Then $G_{cc}(\omega)$ has poles at $\epsilon_{\pm}$ given by 
\begin{equation}
\epsilon_{\pm}=\frac{\left(\epsilon_{a}+\epsilon_{c}\right)}{2}\pm\sqrt{\mathscr{N}\tilde{V}^{2}+\left(\frac{\epsilon_{a}-\epsilon_{c}}{2}\right)^{2}}.\label{Eq.epm}
\end{equation}
 These are the two polaritonic states $|+\rangle$
and $|-\rangle$. The similarity of the above to simple molecular orbital theory \cite{coulson1979}
is to be noted. The residues of these at the poles are easily found
to be 
\[
|\langle c|\pm\rangle|^{2}=\left[1+\frac{\mathscr{N}\tilde{V}^{2}}{(\epsilon_{\pm}-\epsilon_{a})^{2}}\right]^{-1}.
\]
For $\sqrt{\mathscr{N}}|\tilde{V}|\gg|\epsilon_{c}-\epsilon_{a}|$
these reduce to 
\[
\epsilon_{\pm}\simeq\frac{\left(\epsilon_{a}+\epsilon_{c}\right)}{2}\pm|\tilde{V}|\sqrt{\mathscr{N}}
\]
and 
\[
|\langle c|\pm\rangle|^{2}\simeq\frac{1}{2}.
\]
At these poles, the imaginary part $\left[G_{cc}(\omega)\right]^{-1}$
can be easily estimated to be $\approx i\frac{\mathscr{N}\tilde{V}^{2}\pi}{\sqrt{2\pi}\sigma}e^{-\frac{\mathscr{N}\tilde{V}^{2}}{2\sigma^{2}}}$.
This gives the width of the polaritonic peaks. Note that the width of the peak is  decreasing exponentialy with ${\mathscr N}$ and hence its height would increase exponentially.\\

 From Eq. (\ref{Eq.epm}) we see the independence of $\epsilon_{\pm}$
upto first order in $\sigma$.  
To see the effect of disorder, it is necessary to do an analysis up to second order in $\sigma$.
Under the assumption that
the imaginary part is small and keeping the next two terms too in
the asymptotic expansion of $DawsonF(x)$ we can find solutions for
$\epsilon_{\pm}$ up to order $\sigma^{2}$ using MATHEMATICA to be 
\begin{equation}
\epsilon_{s}=\frac{\left(\epsilon_{a}+\epsilon_{c}\right)}{2}+s\sqrt{\mathscr{N}\tilde{V}^{2}+\left(\frac{\epsilon_{a}-\epsilon_{c}}{2}\right)^{2}}+s\frac{\sigma^{2}\left((\frac{\epsilon_{a}-\epsilon_{c}}{2})\left(\frac{\epsilon_{a}-\epsilon_{c}}{2}+s\sqrt{\mathscr{N}\tilde{V}^{2}+(\frac{\epsilon_{a}-\epsilon_{c}}{2})^{2}}\right)+\mathscr{N}\tilde{V}^{2}\right)}{\mathscr{N}\tilde{V}^{2}\sqrt{\mathscr{N}\tilde{V}^{2}+(\frac{\epsilon_{a}-\epsilon_{c}}{2})^{2}}},
\end{equation}
where $s=\pm$. For $\epsilon_{c}=\epsilon_{a}$ this reduces to 
\begin{equation}
\epsilon_{\pm}=\epsilon_{c}\pm\left(\sqrt{\mathscr{N}}|V|+\frac{\sigma^{2}}{\sqrt{\mathscr{N}}|V|}\right).
\end{equation}
 Thus when $\epsilon_{a}=\epsilon_{c}$, the gap between the two polaritonic
states is increased by the disorder, a conclusion which is valid to order $\sigma^{2}$.
Further, as the number of molecules increase, the effects of disorder on the polaritonic states
decrease. These results indicate that the polaritons are quite stable
even in the presence of disorder, provided the number density of molecules
is large enough, i.e., when $\sqrt{\mathscr{N}}|\tilde{V}|\gg\sigma$,
a condition that is satisfied easily. It may be noted that an analysis valid up to first order in $\xi$ was performed in the paper by Houdr$\acute{e}$ \cite{Houdre1996}, who concluded that disorder has no effect on the separation between the polaritonic states. In our notation,  their analysis corresponds to doing calculations correct up to first order in $\sigma$.   From the above it is clear that at this order in $\sigma$, disorder has no effect on the separation between the polaritonic peaks, in agreement with them.

\subsubsection{Testing analytical results using simulations}

We perform numerical calculations to determine the density of states using the equations given in previous sections.
For this we generated ${\mathscr N}$ values of $\xi_{i}$ using a Gaussian
distribution with zero mean and standard deviation $\sigma$. In principle,
we should solve for the roots of $(\omega-\epsilon_{c}-\Sigma_{R}(\omega))$
and then construct the plot. We can easily bypass this by assuming
that $\omega$ has a small imaginary part and then calculating $-\frac{1}{\pi}Im[G_{cc}(\omega)].$
In our calculations, we have taken the imaginary part to be equal to $0.001 eV$. The resulting plot is not very sensitive to the value of this imaginary part. All the other
quantities of interest are evaluated in a similar fashion. For each
quantity we average over 3000 realizations. In Fig.  \ref{Fig3}, \ref{Fig4} and \ref{Fig5} one
can observe the good agreement between the simulations and the theoretical
results even for $\mathscr{N}$ as small as $\mathscr{N}=1500\,m^{-3}$. 
Fig. \ref{Fig3}  shows the close agreement between theoretically calculated $\rho_{c}(\omega)$ with the one obtained from simulations.  In Fig. \ref{Fig4} we show plots of $\rho_T(\omega)$ against $\omega$ for two values of $\sigma$.} Because of the division by $\mathscr{N}$ the polaritonic peaks have a small height and are not clearly visible.   
\begin{figure}[h]
	\includegraphics[width=0.8\linewidth]{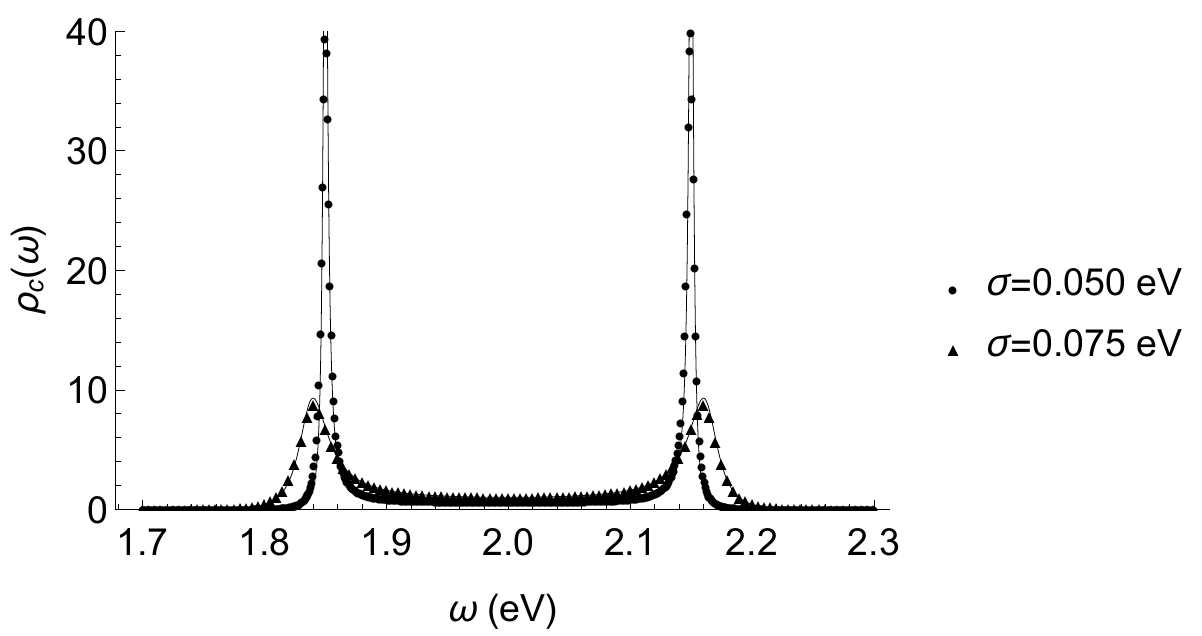} \caption{Plot of $\rho_{c}(\omega)$ against $\omega$ for $\mathscr{N}=1500\,m^{-3}$. Following
		values were used: $\epsilon_{c}=2.0\,eV$, $\epsilon_{a}=2.0\,eV$, $\tilde{V}=3.56\times10^{-3}\,eV m^{3/2}$ and $\eta=0.001\,eV$. The points represents the simulations
		and the line plot is for the analytical  result.}
	\label{Fig3}  
\end{figure}
\begin{figure}[h]
	\includegraphics[width=0.8\linewidth]{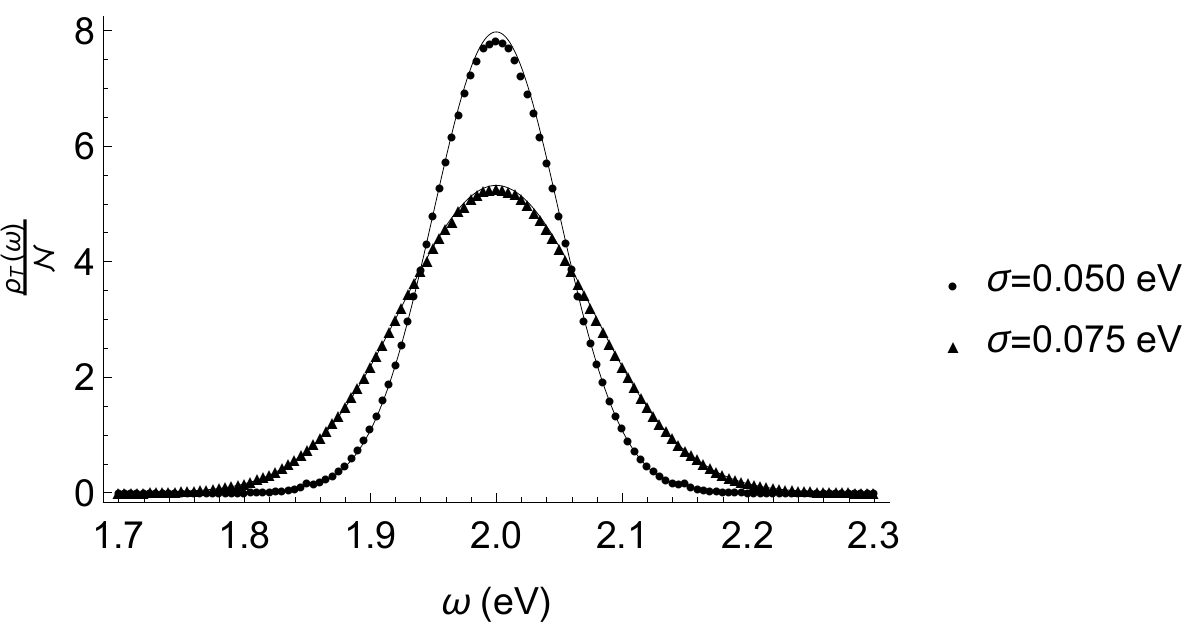} \caption{Plot of $\rho_{T}(\omega)/\mathscr{N}$ against $\omega$ for $\mathscr{N}=1500\,m^{-3}$. Following
		values were used $\epsilon_{c}=2.0\,eV$, $\epsilon_{a}=2.0\,eV$, $\tilde{V}=3.56\times10^{-3}\,eV m^{3/2}$ and $\eta=0.001\,eV$. The points represent the simulations
		and the line plot is for the theoretical result. The polaritonic peaks  barely visible in the plot, due to the division by $\mathscr{N}$.}
	\label{Fig4} 
\end{figure}
Fig. \ref{Fig5} shows the comparison of the analytical and simulated results for  $-\frac{1}{\pi}Im[G_{mol,mol}(\omega)]$.  The agreement between the two is good.
\begin{figure}[h]
	\includegraphics[width=0.8\linewidth]{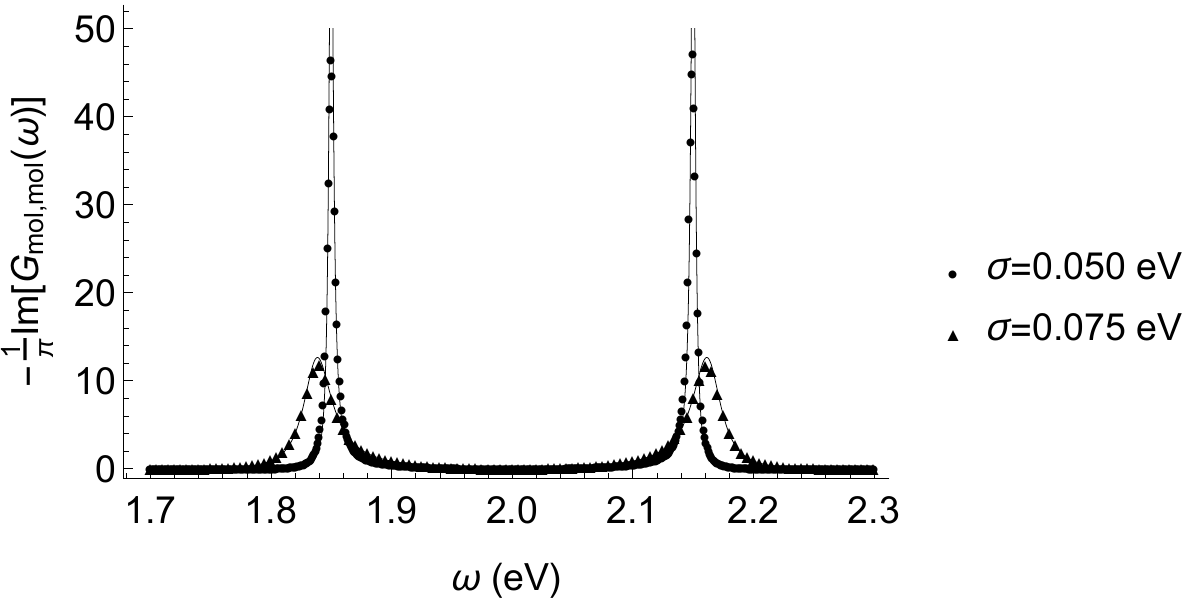} \caption{Plot of $-\frac{1}{\pi}Im[G_{mol,mol}(\omega)]$ against $\omega$ for $\mathscr{N}=1500\,m^{-3}$. Following
		values were used: $\epsilon_{c}=2.0\,eV$, $\epsilon_{a}=2.0\,eV$, $\tilde{V}=3.56\times10^{-3}\,eV m^{3/2}$ and $\eta=0.001\,eV$. The solid points represent the simulations
		and the line plot is for the analytical  result.}
	 \label{Fig5}
\end{figure}

\begin{figure}[h]
	\includegraphics[width=0.8\linewidth]{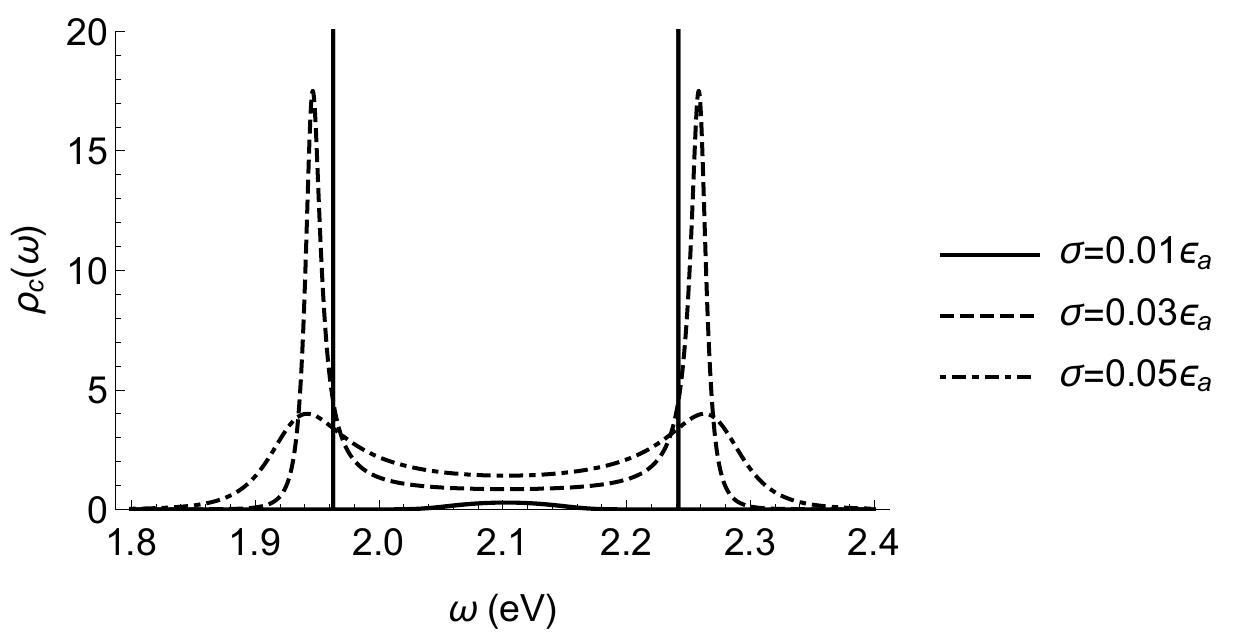} \caption{Plot for $\rho_{c}(\omega)$ v/s $\omega$ for $\mathscr{N}=1.15\times10^{25}\,m^{-3}$ and  for different values of $\sigma$.
		Following values were used: $\epsilon_{c}=2.1\,eV$, $\epsilon_{a}=\epsilon_{c}$,
		$\tilde{V}=4.06\times10^{-14}\,eVm^{3/2}$.}
	\label{Fig6} 
\end{figure}

Having illustrated the results using smaller values  of $\mathscr{N}$, we now extend our results to experimentally viable
values, which are too large to be simulated. As our analysis becomes better and better with increase in $\mathscr{N}$ one expects the analytical results to be accurate for such large values of  $\mathscr{N}$.  In Fig. \ref{Fig6} we plot $\rho_c(\omega)$ against $\omega$ for $\sigma=0.01\epsilon_a,0.03\epsilon_a$ and $0.05\epsilon_a$. The peak heights for the plot corresponding to $\sigma=0.01\,\epsilon_{a}$ value is of the order of $10^{9}$ and hence the top portion has been cutoff. The plots show that the polaritonic states are very much affected by the disorder, as seen from the analytical estimate of their widths which is of the order of $\approx \frac{\mathscr{N}\tilde{V}^{2}\pi}{\sqrt{2\pi}\sigma}e^{-\frac{\mathscr{N}\tilde{V}^{2}}{2\sigma^{2}}}$.  

\begin{figure}[h]
	\includegraphics[width=0.8\linewidth]{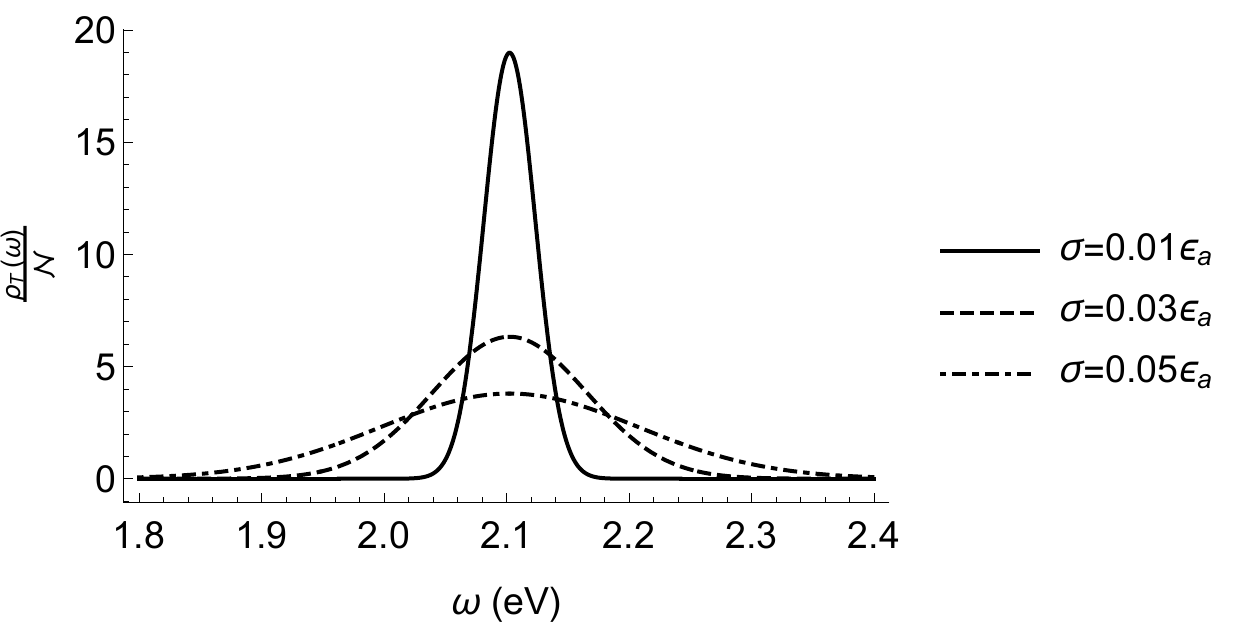} \caption{Plot for $\rho_{T}(\omega)/\mathscr{N}$ v/s $\omega$ for $\mathscr{N}=1.15\times10^{25}\,m^{-3}$ and  for different values of $\sigma$.
		Following values were used: $\epsilon_{c}=2.1\,eV$, $\epsilon_{a}=\epsilon_{c}$,
		$\tilde{V}=4.06\times10^{-14}\,eVm^{3/2}$.}
	\label{Fig7} 
\end{figure}

In Fig. \ref{Fig7} we plot $\rho_{T}(\omega)/\mathscr{N}$ vs $\omega$
for experimentally viable values for different values of $\sigma$.
One does not see the polaritonic peaks in this plot because the height and width of those two peaks are very small for all values of $\sigma$. 

\subsubsection{Change in the density of states}

\begin{figure}[h]
	\includegraphics[width=0.8\linewidth]{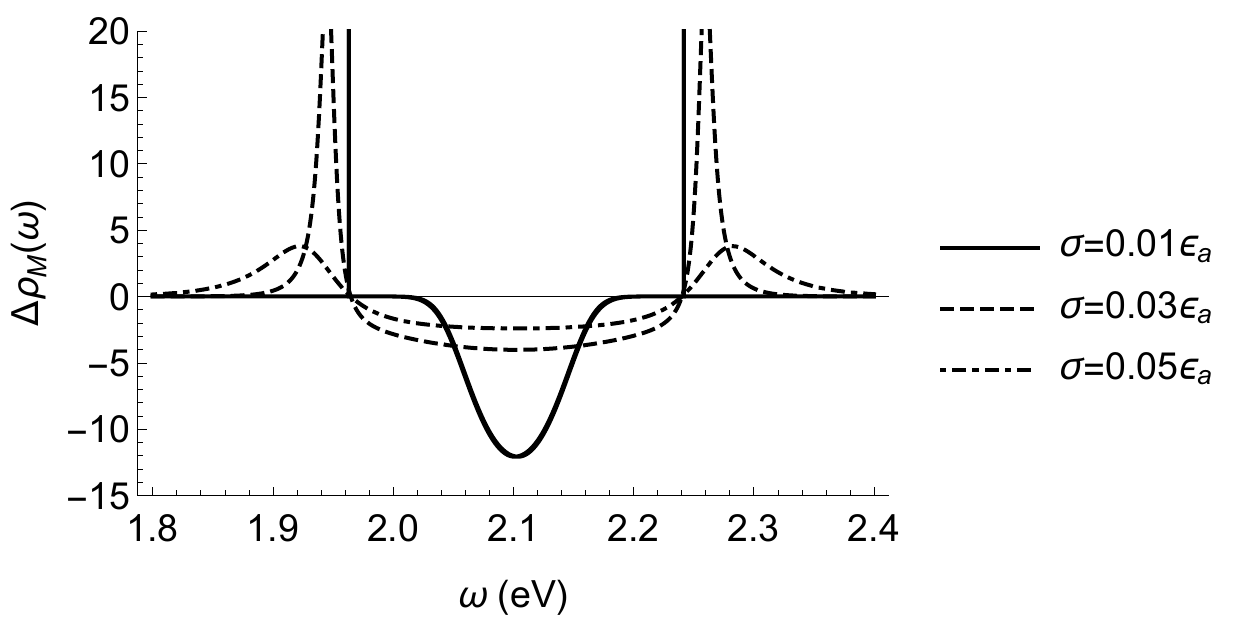} \caption{Plot of change in molecular density of states $\Delta\rho_{M}(\omega)$  vs $\omega$ for $\mathscr{N}=1.16\times10^{25}\,m^{-3}$ and varied values of $\sigma$.
		Following values were used: $\epsilon_{c}=2.1\,eV$, $\epsilon_{a}=\epsilon_{c}$,
		$\tilde{V}=4.06\times10^{-14}\,eVm^{3/2}$. }
	\label{Fig8} 
\end{figure}

Using the expression for the change in density of states given in Eq. (\ref{Eq.drhom}) we plot $\Delta\rho_{M}(\omega)$ in Fig. \ref{Fig8}.
Integrating the area under the curve from $\omega=2.0\; eV$ to $\omega=2.2\; eV$ results in the value of $-1$.
This is what one would expect, because effectively out of the $N$ molecular states
$1$ is used up for the formation of the two polaritonic states. The plot has a minimum at the middle.   This is because  it is the molecular states that are near resonance with the cavity state that contribute the most to the polaritonic states and hence the depletion of density of $\rho_M(\omega)$ is maximum in the middle. In this figure, for $\sigma=0.01\,\epsilon_{a}$ the polaritonic peaks are very high (height  $\approx 10^{9}$) peak with a very narrow (width ~ $\approx 10^{-10}$).    

\subsubsection{Absorption Cross Section}

\begin{figure}[h]
	\includegraphics[width=0.8\linewidth]{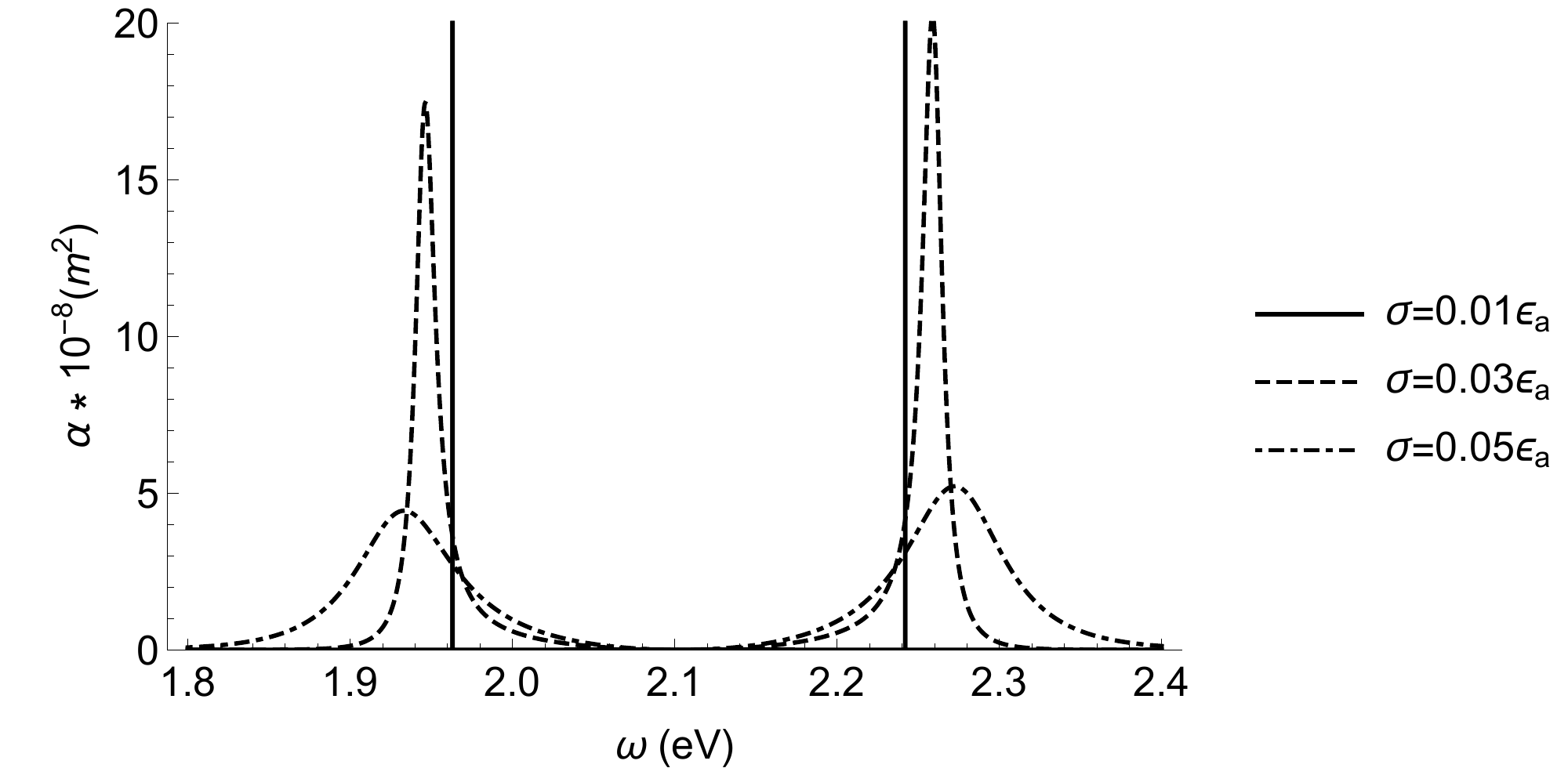} \caption{Plot of $\alpha$ against $\omega$ for $\mathscr{N}=1.15\times10^{25}\,m^{-3}$ with varied values of $\sigma$.
		Following values were used: $\epsilon_{c}=2.1\,eV$, $\epsilon_{a}=\epsilon_{c}$,
		$\tilde{V}=4.06\times10^{-14}\,eVm^{3/2}$ and $|\boldsymbol{\mu}|=10\,D$.   For the sake of clarity of the figure, the y-axis is cutoff at the value of $20$.   The peak height for $\sigma=0.01\epsilon_{a}$ is of the order of $10^{10}$. }  
	\label{Fig9} 
\end{figure}

In Fig. \ref{Fig9} we plot   the absorption cross-section given by Eq. (\ref{Eq.CS}) for increasing values of $\sigma$.  The shift of peak position with increasing values of $\sigma$ is clearly seen. In addition the peaks get broader as $\sigma$ increases. This indicates that with increase in disorder, the dark states are turning ``turning grey". The dependence of width on $\sigma$ is highly non-linear. The approximate relation between $\sigma$ and the width,  determined from the imaginary part of $[G_{cc}(\omega)]^{-1}$ at the poles, is $\approx \frac{\mathscr{N}\tilde{V}^{2}\pi}{\sqrt{2\pi}\sigma}e^{-\frac{\mathscr{N}\tilde{V}^{2}}{2\sigma^{2}}}$. When $\sigma<< 2\sqrt{\mathscr{N}}\tilde{V}$ with $\sigma=0.01\epsilon_{a}=0.02 \,eV$ and $2\sqrt{\mathscr{N}}\tilde{V}=0.275\,eV$, contribution of the dark states to the absorption spectra is miniscule. The absorption is mostly from the two polaritonic states and amounts to $99.86\%$ of the total and the absorption due to the grey states amounts to $0.14\%$ when $\sigma = 0.01\epsilon_{a}$. For $\sigma=0.01\,\epsilon_{a}$, the FWHM for the peaks corresponding to the two  polaritonic states is of the order of $10^{-10}\,eV$ and peak height is of the order of $10^{9}$. As we increase the value of $\sigma$ from $\sigma=0.01 \,eV$ to $\sigma=0.05\epsilon_{a}=0.10 eV$ the contribution of the dark states to the absorption spectra steadily increases.   
\subsubsection{Lifetime effects}
Untill now we considered only the disorder as the source of line broadening. In this section we add the effect of homogeneous broadening for the molecular states $(\gamma_a)$ and the cavity mode linewidth $(\gamma_c)$. For this, we replace $\epsilon_c$ by $\epsilon_c-i\gamma_c$ and $\epsilon_a$ by $\epsilon_a-i\gamma_a$. Then the poles of the $G_{cc}(\omega)$ are at
\begin{equation}
\epsilon_{\pm}=\frac{1}{2} \left(\epsilon_a
+\epsilon_c-i(\gamma_a+\gamma_c)\pm \sqrt{(-\epsilon_a+\epsilon_c+i(\gamma_a-\gamma_c))^2+4
	\mathscr{N}\tilde{V}^2}\right).\end{equation} 
The above expression shows that the polaritons have lifetimes which are appropriately weighted averages of the lifetimes of the cavity and the molecular states, and that for large $\mathscr{N}$, disorder effects on the line-broadening are negligible. 

\begin{figure}[h]
	\includegraphics[width=0.7\linewidth]{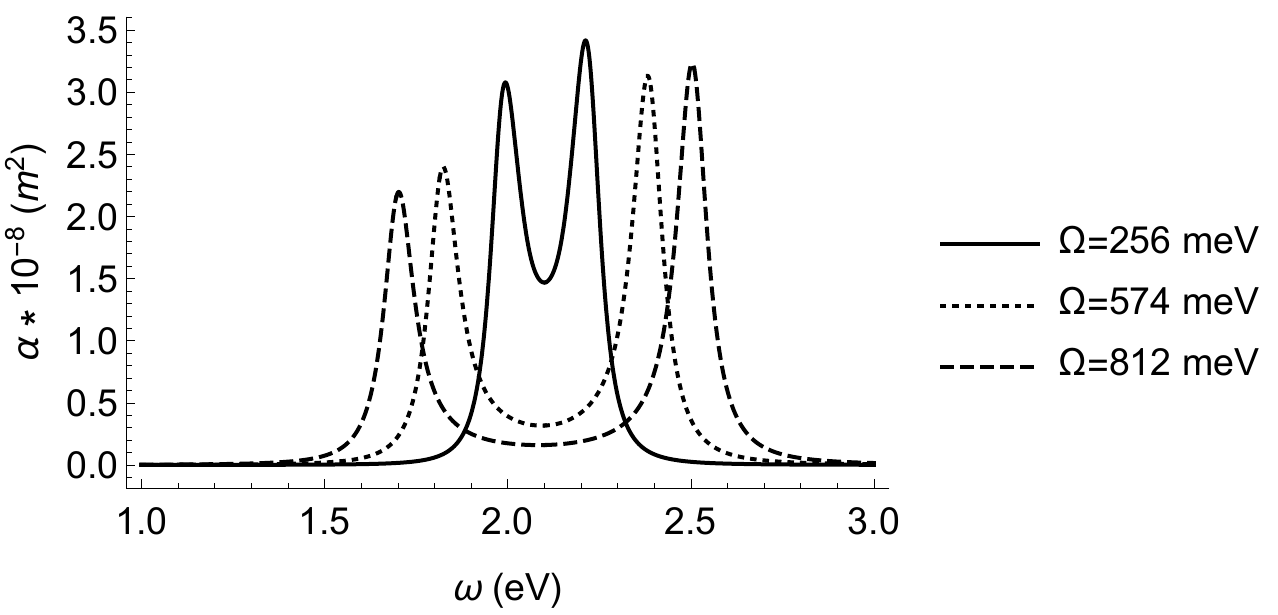} \caption{Plot for absorption cross-section $\alpha$  v/s $\omega$ for varied value of $\Omega $.
		Following values were used: $\epsilon_{c}=2.1\,eV$, $\epsilon_{a}=\epsilon_{c}$,
		$\tilde{V}=4.06\times10^{-14}\,eVm^{3/2}$, $\sigma=0.01\epsilon_{a}$, $\gamma_a=0.3\,meV$ and $\gamma_c=0.1\, eV$.  }
	\label{Fig10} 
\end{figure}

\begin{figure}[h]
	\includegraphics[width=0.7\linewidth]{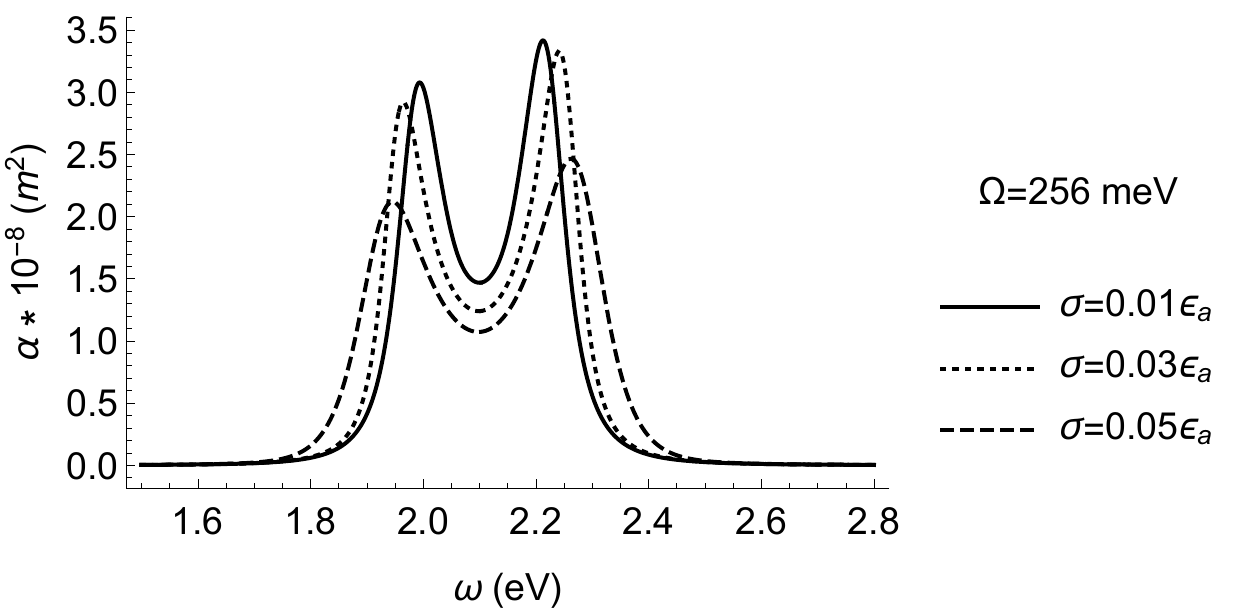} \caption{Plot for absorption cross-section $\alpha$  v/s $\omega$ for varied value of $\sigma $.
		Following values were used: $\epsilon_{c}=2.1\,eV$, $\epsilon_{a}=\epsilon_{c}$,
		$\tilde{V}=4.06\times10^{-14}\,eVm^{3/2}$, $\Omega=256\,meV$, $\gamma_a=0.3\,meV$ and $\gamma_c=0.1\, eV$.}
	\label{Fig11} 
\end{figure}

\begin{figure}[h]
	\includegraphics[width=0.7\linewidth]{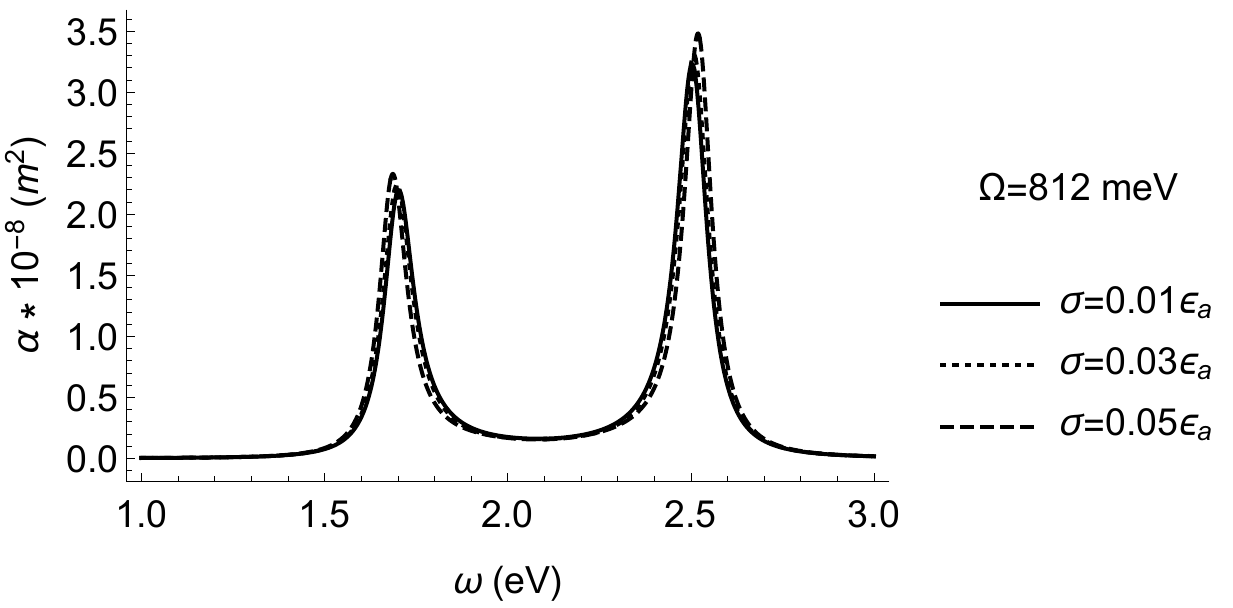} \caption{Plot for absorption cross-section $\alpha$  v/s $\omega$ for varied value of $\sigma $.
		Following values were used: $\epsilon_{c}=2.1\,eV$, $\epsilon_{a}=\epsilon_{c}$,
		$\tilde{V}=4.06\times10^{-14}\,eVm^{3/2}$, $\Omega=812\, meV$, $\gamma_a=0.3\,meV$ and $\gamma_c=0.1\, eV$.}
	\label{Fig12} 
\end{figure}

For typical value of $\gamma_a=10\,ps$ $(0.3\,meV)$ and $\gamma_c=40\,fs$ $(0.1\,eV)$, we study the effect on the lineshape with increasing Rabi-splitting ($\Omega=2\sqrt{\mathscr{N}}\tilde{V}$), shown in Fig. \ref{Fig10}. In the previous section we stated that as $\mathscr{N}$ increases the width of the polaritonic peaks exponentially decreases $\left( \frac{\mathscr{N}\tilde{V}^{2}\pi}{\sqrt{2\pi}\sigma}e^{-\frac{\mathscr{N}\tilde{V}^{2}}{2\sigma^{2}}}\right)$ with it, therefore as the value of $\Omega$ increases, the effects of disorder vanishes exponentially and the contribution of the disorder to the linewidth goes down. In figures \ref{Fig11} and \ref{Fig12} we plot the absorption cross-section with varied values of $\sigma$ keeping $\Omega$ constant. We can observe from Fig. \ref{Fig11} that as $\sigma$ increases the peak position shifts and the lineshape is modified. But as shown in Fig. \ref{Fig12} for a higher value of $\Omega$ there is harldy any effect of $\sigma$ on the lineshape. Therefore when $\epsilon_{a}=\epsilon_{c}$, under the condition $\Omega>>\frac{\gamma_a+\gamma_c}{2}>>\sigma$ the effect of disorder is negligible. This result echoes the findings of Houdr\'e \cite{Houdre1996}.
\section{Model II: $V_{i}=V(\theta_{i},z_{i})$}
\label{Sec:Model2}

The next situation we consider has orientational and position dependent coupling i.e., $V_{i}=V\cos(\theta_{i})\sin(kz_{i}/L)$.
$\theta_{i}$ is the angle between the electric field and transition
dipole moment of the $i^{th}$ molecule. The $\sin(kz_{i}/L)$ term
accounts for the spatial variation of the electric field in the cavity
of length $L$. $\xi_{i}$ is assumed to be identically distributed Gaussian
random variables with the probability distribution same as the one
considered in Model I. As a result, we get\\
\begin{eqnarray}
\langle\Sigma(\omega)\rangle=\frac{V^{2}}{4\pi L}\int_{0}^{\pi}\int_{0}^{L}\int_{0}^{2\pi}d\theta_{i}d\phi_{i}dz_{i}\sum_{i}\frac{\sin(\theta_{i})\cos(\theta_i)^{2}\sin(kz_{i})^{2}}{\omega-\epsilon_{i}}.
\end{eqnarray}
 Averaging over the angles and length will give us: 
\begin{eqnarray}
\langle\Sigma(\omega)\rangle=\frac{V^{2}}{6}\sum_{i}\frac{1}{\omega-\epsilon_{i}}.\label{Eq.AvgSigma}
\end{eqnarray}
This means that effectively the only result of orientational and positional averaging is to reduce the   coupling per
molecule by a factor of $\sqrt{6}$.

\subsection{The Absorption Spectrum for Model II}

Using the general interaction Hamiltonian given by Eq. (\ref{Eq.Hint}),
the absorption cross-section can be written as: 
\begin{eqnarray}
\alpha & = & \frac{\omega}{2\epsilon_{o}c\hbar}\int_{-\infty}^{\infty}e^{i\omega t}\left\langle g\left|\hat{H}_{int}(t)\hat{H}_{int}\right|g\right\rangle dt\label{Eq.CS2-1}
\end{eqnarray}
For this case the definition of $|mol\rangle$ needs to be modified.  Now it is given by 
\begin{equation}
|mol\rangle=F\sum_{i}\boldsymbol{\hat{\mu}}_{i}.\boldsymbol{E}_{i}|g\rangle, \label{Eq.mol}
\end{equation}
where $\boldsymbol{\hat{\mu}}_{i}=\boldsymbol{\mu_i} |g\rangle\langle e_i| + \boldsymbol{\mu_i^{*}} |e_i\rangle\langle g|$
and $F$ is the normalization constant given by: 
\begin{equation}
F^{2}\sum_{i}\left|\boldsymbol{{\mu}}_{i}.\boldsymbol{E}_{i}\right|^{2}=1\label{Eq.F}
\end{equation}
 Using this definition for $|mol\rangle$ we can rewrite Eq. (\ref{Eq.CS2-1}) as
\begin{eqnarray}
\alpha & = & \frac{\omega}{2\epsilon_{o}c\hbar F^{2}}\int_{-\infty}^{\infty}e^{i\omega t}\langle mol\left|e^{-iHt/\hbar}\right|mol\rangle e^{iE_{g}t/\hbar}dt\\
 & = & -\frac{\omega}{\epsilon_{o}c\hbar F^{2}}Im\left\{ G_{mol,mol}(\omega)\right\}. 
\end{eqnarray}
Since $F$ contains the information over orientation and spatial distribution
we substitute it back and average over length and angle, 
\begin{eqnarray}
\alpha & = & -\frac{\omega}{\epsilon_{o}c\hbar}\sum_{i}\left|\boldsymbol{{\mu}}_{i}.\boldsymbol{E}_{i}\right|^{2}Im\left\{ G_{mol,mol}(\omega)\right\} 
\end{eqnarray}
To determine $G_{mol,mol}(\omega)$ we re-write Eq. (\ref{Eq.GM})
\begin{equation}
\underline{\underline{G}}_{M}(\omega^{+})=\left[\omega^{+}-\underline{\underline{\epsilon}}_{M}-\frac{\underline{V}^{\dagger}\underline{V}}{\omega^{+}-\epsilon_{c}}\right]^{-1}.
\end{equation}
Using Eq. (\ref{Eq.mol}) we can write the operator corresonding to
$\underline{V}^{\dagger}\underline{V}$ in the form $F^{-2}|mol\rangle\langle mol|$,
this gives us, 
\begin{equation}
\hat{G}_{M}(\omega^{+})=\left[\omega^{+}-\hat{\epsilon}_{M}-\frac{F^{-2}\left|mol\right\rangle \left\langle mol\right|}{\omega^{+}-\epsilon_{c}}\right]^{-1}.
\end{equation}
Similar to the procedure for model I we can follow the steps to get:
\begin{align*}
{G}_{mol,mol}(\omega^{+})= & \left\{ 1-\frac{F^{-2}\langle mol|(\left[\omega^{+}-\hat{\epsilon}_{M}\right]^{-1})|mol\rangle}{\omega^{+}-\epsilon_{c}}\right\} ^{-1}\langle mol|\left[\omega^{+}-\hat{\epsilon}_{M}\right]^{-1}|mol\rangle\\
= & \left\{ 1-\frac{1}{\omega^{+}-\epsilon_{c}}\sum_{i}\frac{\left|\boldsymbol{{\mu}}_{i}.\boldsymbol{E}_{i}\right|^{2}}{\omega^{+}-\epsilon_{i}}\right\} ^{-1}F^{2}\sum_{i}\frac{\left|\boldsymbol{{\mu}}_{i}.\boldsymbol{E}_{i}\right|^{2}}{\omega^{+}-\epsilon_{i}}
\end{align*}
The average over orientation and length are performed with an approximation that since the terms are all squares of $\left|\boldsymbol{{\mu}}_{eg,i}.\boldsymbol{E}_{i}\right|$,
we can individually average over each term. On averaging,
\begin{align*}
{G}_{mol,mol}(\omega^{+})= & \left\{ 1-\frac{1}{\omega^{+}-\epsilon_{c}}\langle\Sigma(\omega)\rangle\right\} ^{-1}\frac{6}{NV^{2}}\langle\Sigma(\omega)\rangle
\end{align*}
Here, $\langle\Sigma(\omega)\rangle$ is used as given in Eq. (\ref{Eq.AvgSigma}).
With this result the absorption cross-section averaged over the length
and orientation is, 
\begin{eqnarray}
\alpha & = & -\frac{\omega}{6\epsilon_{o}c\hbar}N\left|\boldsymbol{{\mu}}\right|^{2}Im\left\{ G_{mol,mol}(\omega)\right\}. \label{Eq.CS}
\end{eqnarray}
\section{Conclusions}
A general solution for Tavis-Cummings model  for a system with energetic disorder which follows a Gaussian distribution is provided in this work. Analytical solutions are obtained in the limit of large $\mathscr{N}$  for a simple system where coupling is same for all the molecules and for a general system where coupling depends in the position and orientation of the molecules.  Conditions for the existence of the polaritonic states is derived.  A general expression for the polaritonic energies correct upto second order in disorder is derived.  For the case where $\epsilon_a=\epsilon_c$, it is found that increase in disorder leads to an increase in the Rabi splitting.  In general, polaritonic states are found to be very stable against disorder.  They would always exist if the number density of molecules is sufficiently large.  Our calculation of the absorption spectrum shows that  disorder causes the dark states to turn grey. However, in the case where Rabi splitting is sufficiently large, disorder plays almost no role in the line width and in this limit the line width is dominated by lifetime effects. Orientational and position dependence of the coupling is shown to effectively renormalize the coupling strength.
\section{Acknowledgements}
The authors thank Prof. Srihari Keshavamurthy (IITK) for his wonderful lectures titled ``Chemistry with Quantum Light".   This work is a result of an after lecture discussion with him and Prof. Madhav Ranganathan (IITK) and the authors are grateful to both of them.

\end{document}